\documentclass{webofc}

\usepackage[varg]{txfonts}   
\usepackage{hyperref}
\usepackage{url}
\hypersetup{colorlinks=true,citecolor=blue,urlcolor=blue,linkcolor=blue}
\newcommand{\cA}{\vartriangleright}
\newcommand{\cB}{\vartriangleleft}
\newcommand{\dd}{\textrm{d}}

\begin{document}
\title{General Mass treatment for Z boson production in association with a heavy quark at hadron colliders}
%
%

\author{\firstname{M.} \lastname{Guzzi}\inst{1}\fnsep\thanks{\email{mguzzi@kennesaw.edu}} 
\and
            \firstname{P.} \lastname{Nadolsky}\inst{2,3}\fnsep\thanks{\email{nadolsky@smu.edu}}  
        \and
        \firstname{L.} \lastname{Reina}\inst{4}\fnsep\thanks{\email{reina@hep.fsu.edu}}
        \and
        \firstname{D.} \lastname{Wackeroth}\inst{5}\fnsep\thanks{\email{dw24@buffalo.edu}}
        \and
        \firstname{K.} \lastname{Xie}\inst{2}\fnsep\thanks{\email{xiekepi1@msu.edu}}
}

\institute{Department of Physics, Kennesaw State University, 370 Paulding Avenue, Kennesaw, GA 30144, USA
             \and
            Department of Physics and Astronomy, Michigan State University, East Lansing, MI 48824, USA 
            \and
            Department of Physics, Southern Methodist University, Dallas, TX 75275-0181, USA
            \and
           Physics Department, Florida State University, Tallahassee, FL 32306-4350, USA
           \and
           Department of Physics, University at Buffalo, The State University of New York, Buffalo, New York 14260-1500, USA
          }

\abstract{We present the application of the ACOT and S-ACOT general mass variable flavor number schemes to proton-proton collisions with particular attention to the production of final states with at least one heavy quark. \textit{Subtraction} and \textit{residual} heavy-quark parton distribution functions are introduced to facilitate the implementation of this scheme at higher orders in perturbative QCD.  The calculation of $Z$-boson hadronic production with at least one $b$ jet beyond the lowest order in QCD is considered for illustration purposes.
}
\maketitle

\section{Introduction}
\label{intro}
Modern global QCD analyses aimed at determining the proton's parton distribution functions (PDFs) and the substantial influx of high-precision measurements from the Large Hadron Collider (LHC), necessitate precise and accurate theoretical predictions for standard-candle processes in perturbation theory. The magnitude of higher-order radiative corrections in perturbative QCD calculations competes with multiple factors. Heavy-quark (HQ) contributions can play a significant role among those, and their dynamics can influence the cross-section calculations for a wide range of crucial processes at current and future colliders.
Global PDF analyses at next-to-next-to-leading order (NNLO) and beyond (see refs.~\cite{McGowan:2022nag,NNPDF:2024nan}) extend on a wide range of collision energies and are sensitive to mass effects such as phase-space (PS) suppression, and large radiative corrections to collinear $Q \bar Q$ production. The magnitude of these effects is comparable to NNLO, and next-to-next-to-next-to-leading order (N$^3$LO) corrections. With the recent progress in theory calculations beyond NNLO in the QCD strong coupling (e.g., N$^3$LO DIS ~\cite{Vermaseren:2005qc,Moch:2004xu,Moch:2007rq,Moch:2008fj,Kawamura:2012cr,Davies:2016ruz,Blumlein:2022gpp}, DIS with massive quarks~\cite{Ablinger:2024qxg}, N$^3$LO Drell-Yan~\cite{Baglio:2022wzu,Duhr:2020sdp,Duhr:2021vwj,Chen:2021vtu,Chen:2022lwc}, N$^3$LO Higgs production~\cite{Anastasiou:2015vya,Mistlberger:2018etf,Dulat:2018bfe,Dreyer:2016oyx,Cieri:2018oms}, and jet production in DIS~\cite{Gehrmann:2018odt,Currie:2018fgr}), HQ contributions must be consistently considered in the cross section calculation for particle reactions to advance the precision frontier of the LHC Run-3 and beyond.  
That is, all the fitted cross sections should be computed in a factorization scheme able to interpolate between calculations in which the mass is fully retained  in the appropriate kinematic regimes (Fixed Flavor Number (FFN) scheme), and calculations where the number of active flavors $N_f$ increases with the typical energy scale and where mass effects may be approximated when the HQ mass is negligible compared to the other scales (e.g., Zero-Mass (ZM) variable flavor number (VFN) scheme). This matching must guarantee the continuity of the theoretical predictions and avoid double counting of scattering contributions. These interpolating schemes are generically referred to as General Mass Variable Flavor Number (GMVFN) schemes.
In this work, we apply the Aivazis-Collins-Olness-Tung (ACOT) GMVFN scheme~\cite{Aivazis:1993kh,Aivazis:1993pi}, and its simplified version (S-ACOT)~\cite{Tung:2001mv,Kramer:2000hn,Guzzi:2011ew} to the production of a $Z$ boson with at least one heavy quark in proton-proton collisions~\cite{Guzzi:2024can}. 
Various versions of the ACOT/S-ACOT schemes have been used to effectively describe heavy-flavor dynamics in DIS. For instance, the S-ACOT-$\chi$ at NNLO for neutral~\cite{Guzzi:2011ew} and charged currents~\cite{Gao:2021fle}, and the H-VFN scheme~\cite{Kusina:2013slm} have been applied. Similar schemes have also been used in inclusive charm/bottom production in proton-proton collisions, such as the S-ACOT-Massive-Phase-Space(-MPS) scheme~\cite{Xie:2019eoe,Xie:2021ycd,Xie:2022sqa}, and in $D$-meson hadroproduction, like the S-ACOT-$m_T$~\cite{Helenius:2018uul}.
The concept of \textit{subtraction} and \textit{residual} HQ PDFs is introduced to facilitate the practical implementation of the ACOT scheme for $Z+b$ production. 
These are expressed as convolutions between PDFs and universal operator matrix elements (OMEs) which represent the transition from a massless parton to a HQ and include the HQ mass dependence. To allow for fast calculations, subtraction PDFs are pre-calculated and provided to external users as part of the CT18~\cite{Hou:2019efy} family of PDFs with grids in the LHAPDF6 format~\cite{Buckley:2014ana} and are publicly available through the HEPForge repository~\cite{sacotmps}. We illustrate how subtraction PDFs are incorporated in the calculation of the hadronic production of a $Z$-boson in association with at least one $b$-quark jet beyond the lowest-order in QCD, using theory predictions for the differential cross section from previous work~\cite{FebresCordero:2008ci,FebresCordero:2009xzo,Figueroa:2018chn}.

\section{GMVFN hadronic cross section for $Z+b$ production at NLO in QCD}
\label{sec-1}

The general structure of a GMVFN scheme hadronic cross section calculation can schematically be represented by the combination of three terms 
\begin{equation}\label{eq:VFN}
\dd\sigma_{\rm GMVFN}=\dd\sigma_{\rm FC}+\dd\sigma_{\rm FE}-\dd\sigma_{\rm Sub} 
\end{equation}
where $\dd\sigma_{\rm FC}$ represents the Flavor Creation (FC) contributions which contain HQs only in the final state and are computed with $m_Q\neq 0$, $\dd\sigma_{\rm FE}$ represents the Flavor Excitation (FE) terms which are HQ-initiated processes where the large collinear logarithms have been conveniently resummed into a HQ PDF, and $\dd\sigma_{\rm Sub}$ is the subtraction which incorporates terms that remove the overlapping part between the FC and FE terms to avoid double counting. 
The implementation of Eq.~(\ref{eq:VFN}) in the ACOT scheme for $Z+b$ production in proton-proton collisions at NLO in QCD requires that the FC contributions represented by the subprocess $i+j\rightarrow Z+b\bar b+X$ with $ij=\{gg,q \bar q, qg, {\bar q}g\}$, where $X$ is any additional radiated light parton, are matched to the FE terms $i+j\rightarrow Z+b+X$ where $ij=\{Qg, \bar Q g, qQ, {\bar q}Q, {\bar Q}q\}$  at ${\cal O}(\alpha_s^3)$.  
The partonic channels contributing to $\dd\sigma_{\rm GMVFN}^{\rm NLO}$ can explicitly be given in terms of PDF \textit{subtractions} as follows (see Ref.~\cite{Guzzi:2024can} for more details on the general framework)
\begin{equation}
\begin{aligned}
\dd\sigma_{\rm FC}^{\rm NLO} &=
f_{g} \cA \left[a_s^2 ~\dd\widehat{\sigma}^{(2)}_{gg\to ZQ\bar{Q}} + a_s^3 ~\dd\widehat{\sigma}^{(3)}_{gg\to ZQ\bar{Q}(g)}  \right]\cB f_{g}
\\
&+\sum_{i=q, \bar q}f_{i}\cA\left[a_s^2~\dd\widehat{\sigma}^{(2)}_{q\bar{q}\to ZQ\bar{Q}}+ a_s^3~\dd\widehat{\sigma}^{(3)}_{q\bar{q}\to ZQ\bar{Q}(g)}\right] \cB f_{\bar i} \\
&+a_s^3\sum_{i=q,\bar q} \left[  f_{g} \cA  \dd\widehat{\sigma}^{(3)}_{gq\to ZQ\bar{Q}(q)} \cB f_{i} 
+ f_{i} \cA \dd\widehat{\sigma}^{(3)}_{qg\to ZQ\bar{Q}(q)} \cB f_{g}\right];
\label{eq:tot-FC}
\end{aligned}\end{equation}
\begin{equation}\begin{aligned}
\dd\sigma_{\rm FE}^{\rm NLO}
&=f_{g}\cA  \left[a_s~\dd\widehat{\sigma}^{(1)}_{gQ\to ZQ}+
a_s^2~\dd \widehat{\sigma}^{(2)}_{gQ\to ZQ(g)}\right]\cB f_{Q}
\\
&+ a_s^2 \sum_{i=q,\bar q} f_i \cA \dd\widehat{\sigma}^{(2)}_{qQ\to ZQ(q)}\cB f_{Q}
+(\rm{exch.});
\label{eq:tot-FE}
\end{aligned}\end{equation}
\begin{equation}\begin{aligned}
\dd\sigma_{\rm sub}^{\rm NLO}
&=a_s~f_g\cA \dd\widehat{\sigma}^{(1)}_{gQ\to ZQ}\cB \tilde{f}_Q^{(\rm NLO)}
+a_s^2 ~f_g \cA 
\dd\widehat{\sigma}^{(2)}_{gQ\to ZQ(g)}\cB \tilde{f}_Q^{(1)}
\\
&+ a_s^2 \sum_{i=q,\bar q} f_i \cA  
\dd\widehat{\sigma}^{(2)}_{qQ\to ZQ(q)}\cB \tilde f_Q^{(1)}
+(\rm{exch.})\,,
\label{eq:tot-subtraction-pdf}
\end{aligned}\end{equation}
where $a_s\equiv \alpha_s(\mu,N_f)/(4\pi)$ and $Q$ is the $b$-quark. The caret symbol `` $\widehat{}$ '' in Eqns.~(\ref{eq:tot-FC}-\ref{eq:tot-subtraction-pdf}) denotes infrared-safe contributions (i.e., dimensional regularization is applied for initial-state massless partons and the associated $1/\epsilon$ singularities are subtracted). Besides the HQ mass and other kinematic variables, $\dd\widehat{\sigma}_{ij}$ depends on the two longitudinal momentum fractions (e.g., $x_A$ and $x_B$) of the initial-state partons. Therefore, to make the implementation more transparent, we introduce the triangular arrows $\cA$ and $\cB$ to define convolution products with two variables where the triangular arrows point to the corresponding integration variable that is integrated in the convolution
\begin{align*}
\label{eq:cAB}
\left[f\cA H\right](x_A, x_B) \equiv \int_{x_A}^{1} \frac{\dd\xi_A}{\xi_A} f(\xi_A) H(\widehat{x}_A, x_B)\,;
\hspace{0.5cm}
\left[H\cB f\right](x_A, x_B) \equiv \int_{x_B}^{1} \frac{\dd\xi_B}{\xi_B} H(x_A,\widehat{x}_B) f(\xi_B).
\end{align*}
With one variable, we then have
\begin{equation}
\begin{aligned}
\label{eq:otimes}
\int_{x}^{1} \frac{\dd\xi}{\xi} f(\xi) g\left(\frac{x}{\xi}\right) = \left[f\cA g\right](x) = \left[g\cB f\right](x).
\end{aligned}
\end{equation}
Finally, the \textit{subtraction} HQ PDF  $\tilde{f}_Q$ in Eq.~(\ref{eq:tot-subtraction-pdf}), in its first two orders in $a_s$, is defined as~\cite{Guzzi:2024can,Xie:2019eoe,Xie:2021ycd}
\begin{equation}
\begin{aligned}
&\tilde{f}_Q^{(1)} = a_s [A^{(1)}_{Qg} \cB g],~
\tilde{f}^{(2)}_Q = a_s^2 \sum_{i=g, q,\bar q} [A^{(2)}_{Qi} \cB f_i]\,,
\end{aligned}
\label{eq:sub-PDFs-def}
\end{equation}
and the NLO subtraction PDF is\footnote{Here ``NLO'' refers to the $a_s$ order of the subtraction coefficients. }
\begin{equation}\begin{aligned}
\tilde{f}_Q^{\rm(NLO)}(x,\mu) \equiv \tilde{f}_Q^{(1)}+ \tilde{f}_Q^{(2)}=a_s [ A^{(1)}_{Qg} \cB g](x,\mu) + a_s^2 \sum_{i=g,q,\bar q} [A^{(2)}_{Qi} \cB f_i](x,\mu)\,,  
\label{eq:sub-PDFs-NLO}
\end{aligned}\end{equation}
where the $A_{ij}^{(n)}$ $(n=1,2,\dots)$ are perturbative coefficients obtained from the calculation of operator matrix elements (OMEs) 
that emerge from mass factorization~\cite{Witten:1975bh,Buza:1995ie,Buza:1996wv}. 
The difference $\dd\sigma_{\rm FE}-\dd\sigma_{\rm sub}$ can equivalently be reorganized in terms of HQ PDF \textit{residuals} as follows
\begin{equation}\begin{aligned}
\dd\sigma_{\rm GMVFN}^{\rm NLO}
&= d\sigma_{\rm FC}^{\rm NLO} + a_s~f_{g}\cA\left[\dd\widehat{\sigma}^{(1)}_{gQ\to ZQ}\right]\cB \delta f_{Q}^{(\mathrm{NLO})} 
\\
&+ a_s^2~f_{g} \cA \left[ \dd\widehat{\sigma}^{(2)}_{gQ\to ZQ(g)}\right]\cB \delta f_{Q}^{(1)}
+a_s^2~\sum_{i=q,\bar q} f_i \cA \left[\dd\widehat{\sigma}^{(2)}_{qQ\to ZQ(q)}\right]\cB \delta{f}^{(1)}_Q
+(\rm{exch.})\,.
\end{aligned}
\label{eq:tot-residual-pdf}
\end{equation}
The quark-antiquark annihilation channel $q\bar{q}\to ZQ\bar{Q}$ does not contain FE terms and therefore does not involve subtraction contributions.  
The realization in Eq.~(\ref{eq:tot-residual-pdf}) includes the minimal set of PQCD contributions required by an ACOT-like scheme at NLO. It can be further streamlined by including more higher-order contributions, see~\ref{Guzzi:2024can}.

\section{Results}

\begin{figure}
\includegraphics[width=0.49\textwidth]{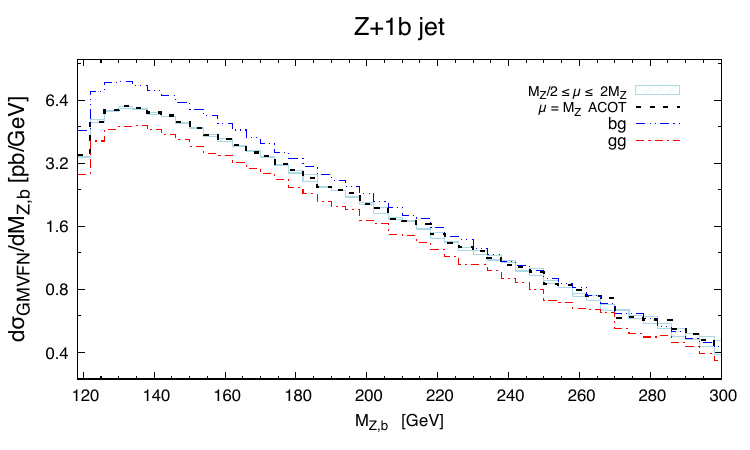}
\includegraphics[width=0.49\textwidth]{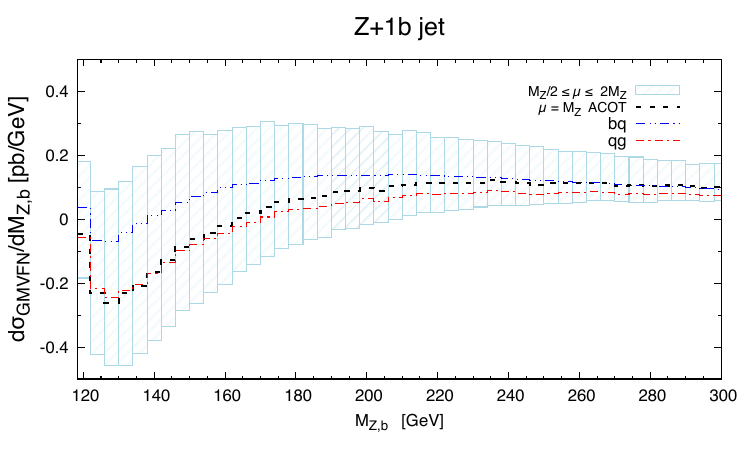}
\caption{$M_{Z,b}$ differential distribution with scale dependence $M_Z/2<\mu<2M_Z$ at order ${\cal O}(\alpha_s^3)$ for the ACOT GMVFN scheme with FC and FE terms in the $gg$(left) and $qg$(right) channels. The CT18NLO PDFs are used.}
\label{MZ-NLO}
\end{figure}

\begin{figure}
\includegraphics[width=0.9\textwidth]{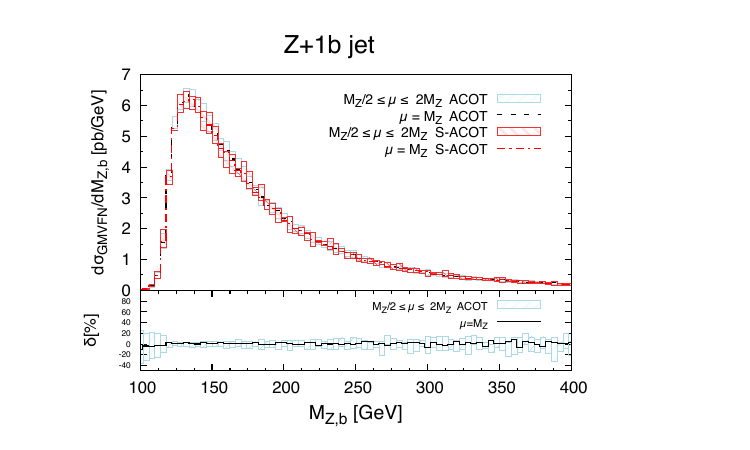}
\caption{$M_{Z,b}$ differential distribution with scale dependence $M_Z/2<\mu<2M_Z$ at order ${\cal O}(\alpha_s^3)$ for the ACOT and S-ACOT GMVFN schemes. The CT18NLO PDFs are used.}
\label{acot-vs-sacot:MZb-NLO}
\end{figure}

The NLO ACOT theory prediction for the invariant mass distribution of the $Z+b$ system $M_{Z,b}$ is illustrated in Fig.~\ref{MZ-NLO} where shown are also the scale uncertainty and the interplay between FC and FE contributions in the $gg$ and $qg$ channels. A more detailed discussion with results for other distributions is in Ref.~\cite{Guzzi:2024can}. The ACOT-$gg$($qg$) central prediction is represented by a black dashed line, while its scale dependence $M_Z/2\leq \mu\leq 2 M_Z$ is represented by a light-blue band. The FC and FE terms are represented by a red dot-dashed and a blue dot-dot-dashed line, respectively.
We observe that the convergence appears to be faster in the $qg$ channel, where the FE cross section is largely cancelled by the subtraction terms at $M_{Z,b}\approx 120$ GeV in contrast to the $gg$ channel, where the matching of terms seems to be more sensitive to phase-space integration and applied cuts.
In Fig.~\ref{acot-vs-sacot:MZb-NLO} we show a comparison between the ACOT and S-ACOT theory predictions within their corresponding scale uncertainty bands (light-blue for ACOT and light-red for S-ACOT). $\delta[\%]$ in the lower inset, represents the percent difference between S-ACOT and ACOT relative to ACOT, within the scale uncertainty of the ACOT prediction. Differences are in general around 2-3\% and smaller, but they can be larger at higher values of the hard scale of the process. 
However, all differences are well within the ACOT scale dependence.    


\section{Conclusions}

We extended the framework of the ACOT/S-ACOT GMVFN scheme, based on QCD factorization to the case of proton-proton collisions where at least one heavy quark is produced. Such a GMVFN scheme is applied to $Z$ boson production in association with a least one $b$-quark jet at ${\cal O}(\alpha_s^3)$ (NLO) in QCD as an illustrative case, and results are presented for the invariant mass distributions of the $Z+b$ system at the LHC with a collision energy of $\sqrt{s}=13$ TeV.
The practical realization of the theory calculation for the $Z+b$ cross section is facilitated by the introduction of subtraction and residual HQ PDFs.
Tabulated grids for CT18 NLO and CT18 NNLO PDF ensembles, in which the $b$-quark PDFs are replaced by either the residual or subtraction $b$-quark PDFs, are provided in the LHAPDF6 format~\cite{Buckley:2014ana} and distributed through a repository at HEPForge~\cite{sacotmps}. These grids are process-independent and can be applied to construct ACOT-like GMVFN theory predictions for other processes.

{\bf Acknowledgements.}
The work of L.R. is supported in part by the U.S. Department of Energy under grant DE- SC0010102. M.G. is partially supported by the National Science Foundation under Grants No.~PHY-2112025 and No.~PHY-2412071.
P.M.N. was partially supported by the U.S. Department of Energy under Grant No.~DE-SC0010129. The work of D.W. is supported in part by the National Science Foundation under Grants No.~PHY-2014021 and No.~PHY-2310363. K.X. is supported by the U.S. National Science Foundation under Grants No.~PHY-2310291 and PHY-2310497.

%
%
%

\bibliographystyle{woc}

\end{document}